# The Crowd Classification Problem:
# Social Dynamics of Binary Choice Accuracy

Joshua Becker[1]

Douglas Guilbeault[2]

Ned Smith[3]

[1] UCL School of Management, University College London

[2] Haas School of Business, University of California Berkeley

[3] Kellogg School of Management, Northwestern University

Correspondence to:

joshua.becker@ucl.ac.uk

**The Crowd Classification Problem: Social Dynamics of Binary Choice Accuracy**

Decades of research suggest that information exchange in groups and organizations can reliably improve judgment accuracy in tasks such as financial forecasting, market research, and medical decision-making. However, we show that improving the accuracy of numeric estimates does not necessarily improve the accuracy of decisions. For binary choice judgments, also known as classification tasks—e.g. yes/no or build/buy decisions—social influence is most likely to grow the majority vote share, regardless of the accuracy of that opinion. As a result, initially inaccurate groups become increasingly inaccurate after information exchange even as they signal stronger support. We term this dynamic the "crowd classification problem." Using both a novel dataset as well as a reanalysis of three previous datasets, we study this process in two types of information exchange: (1) when people share votes only, and (2) when people form and exchange numeric estimates prior to voting. Surprisingly, when people exchange numeric estimates prior to voting, the binary choice vote can become less accurate even as the average numeric estimate becomes more accurate. Our findings recommend against voting as a form of decision-making when groups are optimizing for accuracy. For those cases where voting is required, we discuss strategies for managing communication to avoid the crowd classification problem. We close with a discussion of how our results contribute to a broader contingency theory of collective intelligence.

**1. Introduction**

Numeric estimates (e.g. forecasts) are important in organizational decision-making, such as strategic decisions informed by economic and industry forecasting as well as operational decisions informed by quantities such as project development cost. While such decisions are often made by the expert judgement of a single manager, both statistical and empirical findings show that the aggregated opinion of multiple people can be more accurate than even the most expert individuals (Ashton 1986, Hogarth 1978, Nofer and Hinz 2014), a phenomenon known as the "wisdom of crowds" (Atanasov et al. 2017, Budescu and Chen 2014, Da and Huang 2020, Frey and van de Rijt 2020, Mannes 2009, Palley and Soll 2019).

A central practical question is whether and how social influence between contributors impacts the accuracy of the contributed estimates (Da and Huang 2020, Dalkey and Helmer 1963, Frey and van de Rijt 2020, Minson et al. 2018, Ven and Delbecq 1974). Taken together, current research suggests that the manager faced with a forecasting or estimation task has a simple strategy: ask contributors to first generate an independent estimate (Minson et al. 2018), then discuss in a way that lets everyone participate equally (Becker et al. 2017, 2020, Golub and Jackson 2010), and finally use the resulting group opinion.

However, these prior studies have shown only that social influence increases the accuracy of the resulting numeric estimates—and a number is not in itself a decision. In practice, numeric estimates must be converted to a decision as when a group considering the expected return of an investment takes a vote on whether or not to make the deal. For individuals, converting a numeric estimate to a decision is straightforward: the investment either offers a positive return, or it does not. For groups, however, you can get a different recommendation by taking an average than you would by taking a vote (Csaszar and Eggers 2013, Hastie and Kameda 2005). We study how social information exchange impacts the accuracy of such binary choice votes.

As an illustration, consider a hypothetical manager facing a software "build or buy" decision. Faced with the need for a software solution (e.g., a new client database) this manager can either purchase an existing solution for a known cost or employ a team of developers to build something new for an uncertain cost. Estimating software development cost is a commonly studied forecasting problem that in practice is frequently accomplished with subjective "gut" judgements by managers (Jørgensen 2004). The decision itself comes down to a simple determination of which one is more expensive, an example of a more general class of decisions made by comparing numeric estimates to a benchmark threshold.

More complex decisions may involve multiple uncertain quantities, as in a market entry decision based on estimates for consumer demand, project development cost, and conversion rate for existing customers. Such complex decisions are often reducible to a single summary statistic, e.g. the expected payoff in rational choice models, as recommended by prescriptive theories of managerial decision-making (Bazerman and Moore 1994, Medvec and Galinsky 2005). When negotiating any deal, for example, the decision ultimately reduces to whether the offer on the table is better than the best alternative (Bazerman et al. 2000). In medical decision-making, multiple factors such as estimated cost and expected benefit are reduced to a single metric, and treatment may be given if that metric exceeds a benchmark threshold (Eichler et al. 2004).

Critically, we find that the same social processes which increase the accuracy of numeric estimates can simultaneously decrease the accuracy of the resulting binary choice vote. This dynamic is possible because the numeric estimate and binary vote can become decoupled during social exchange, moving in opposite directions as the result of group information processing. Because binary choice decisions are known as 'classification tasks' in statistical estimation theory (machine learning) we term this dynamic the 'crowd classification problem.'

Briefly stated, the crowd classification problem manifests as the tendency for the initial majority vote share to grow—regardless of accuracy. We first consider the simple case where people communicate their vote only (henceforth "binary exchange"). These dynamics can be summarized with the simple observation that a person in the majority is unlikely to change their vote—only people with a minority opinion flip, and so the majority grows.

However, communicating only binary votes eliminates the nuance carried in a numeric estimate. Thus one interpretation is that the problem arises from communication style. Based on the prior finding that social influence improves numeric estimates, we consider a possible solution: encourage group members to form and share numeric estimates prior to taking a vote (henceforth "numeric exchange"). Even in this case, however, the majority is often amplified regardless of accuracy. Surprisingly, we find that the binary vote can become less accurate even as the numeric estimate becomes more accurate. Figure 1 in §2 below shows an example of how this can occur.

The rest of this paper proceeds as follows. First, we review prior literature on belief accuracy and social influence. Then, we explain the dynamics of the crowd classification problem on an intuitive level. Next, we provide a formal theoretical analysis of both types of communication (binary exchange and numeric exchange). Finally, we present empirical evidence for the crowd classification problem, first in binary exchange and then in numeric exchange.

We close by discussing practical implications and possible solutions to the crowd classification problem. These conclusions can be summarized with a brief rule of thumb: for groups that must make decisions by voting, accuracy is optimized by independence; for groups where alternative decision processes are possible, accuracy can be optimized via social learning.

### 1.1. Belief Accuracy and Social Influence

One of the most widely studied estimation tasks is forecasting (e.g. Atanasov et al. 2017, Dalkey and Helmer 1963, Jansen et al. 2016) including forecasting revenue (Da and Huang 2020) and sales (Cowgill & Zitzewitz, 2015), predicting the success of an advertising campaign (Hartnett, Kennedy, Sharp, & Greenacre, 2016), or estimating future macroeconomic indicators (Jansen, Jin, & de Winter, 2016). In one classic case study, Cyert and March (1963) describe a construction firm for whom expectations of future business volume played a central role in the decision to move operations to a new location. While forecast accuracy is a widely studied area of belief accuracy, the principles identified in this article broadly inform judgements based on the estimation of any uncertain quantity, i.e. "nowcasting" (Jansen et al. 2016).

Numeric estimates can play a role even in decisions where there is no obviously "correct" choice, as in hiring decisions. Hiring ideally involves predicting factors such as performance, cultural fit, and likelihood of remaining employed (Arlotto et al. 2014, Rivera 2012). While many important employee characteristics are subjective, employees at lower levels of organizations are frequently evaluated with explicitly quantified metrics. Industries such as customer service call centers have a long history of quantifying employee productivity with metrics such as quality assurance scores and upsell rates (Holman et al. 2002). More recent advances in monitoring technology are allowing performance quantification for employees as varied as restaurant waitstaff (Pierce et al. 2015) and warehouse workers (Moore 2019). Thus a hiring decision becomes a forecasting task—what score will that employee receive?

When crowdsourcing numeric estimates, it has been popularly argued (Surowiecki 2004) that group accuracy requires strictly independent individuals, and that interacting groups are subject to risks associated with "herding" (Lorenz et al. 2011) and "correlated error" (Hong et al. 2016). Thus one area of research has been the development of strategies designed to optimize the aggregation of beliefs from a multiple independent contributors (Atanasov et al. 2017, Budescu and Chen 2014, Da and Huang 2020, Mannes et al. 2014, Palley and Soll 2019). This paradigm is motivated in part by the "diversity prediction theorem" by Page (2007), a reinterpretation of the variance-bias tradeoff in statistical estimation which states the following[1]: group error = average individual error – diversity. In this equation, group error is the error of the average estimate and diversity is the variance of individual estimates.

A common interpretation of the diversity prediction theorem is that a decrease in diversity leads to an increase in the group error. This interpretation is motivated by the algebraic observation that a decrease in the rightmost term in the equation (diversity) must accompany an increase in the left term of the equation (group error). Notably, however, this only holds true if the middle term (average individual error) remains constant.

However, individual error may not remain constant, and this equation also highlights a clear potential benefit of social influence. Suppose that after discussion, every group member holds an estimate equal to the pre-discussion average. Then final variance equals zero. While this outcome may be interpreted as reduced diversity, this outcome also means that average individual error after discussion is equal to the group error, and that group error itself remains the same. Since group error is guaranteed to be lower than individual error when variance/diversity is non-zero (i.e., before discussion), this convergence-to-the-mean process must necessarily decrease individual error. Empirically, the benefit of social influence for individuals in a group has been observed even for studies that were designed to show the risk of social influence (see letter by Farrell, 2011, in response to Lorenz et al., 2011).

Moreover, a growing body of research has shown that under a wide range of conditions, social influence can even reduce group error. Early research in this direction motivated the development of the "Delphi method" (Dalkey and Helmer 1963) which used a carefully mediated process (people communicated via slips of paper) to allow social learning while mitigating the risks of conformity pressure (Asch 1951). While this early research often produced conflicting results (Hastie 1986), more recent research has begun to identify when social influence will and won't improve belief accuracy by founding experimental design on formal statistical and agent-based models (Almaatouq, Rahimian, et al. 2020, Becker et al. 2020).

---

[1] Formally: $(\bar{x} - \theta)^2 = \frac{1}{N}\sum_{i=1}^{N}(x_i - \theta)^2 - \frac{1}{N}\sum_{i=1}^{N}(x_i - \bar{x})^2$ where an individual $i$ in a population of $N$ people holds belief $x_i$ and the true value is $\theta$. While this applies to squared error, a similar result applies to mean absolute deviation. In general, this will hold for any convex error function.

A general principle is that when everyone can participate equally as determined by factors such as discussion dynamics (Becker et al. 2020) and network structure (Becker et al., 2017), group information exchange is likely to improve belief accuracy as long as people begin with independently formed opinions (Minson et al. 2018). The potential benefit of information exchange among people making numeric estimates has been corroborated by multiple experiments (Almaatouq et al. 2020, Atanasov et al. 2017, Becker et al. 2017, Farrell 2011, Gürçay et al. 2015, Jayles et al. 2017, Minson et al. 2018, Navajas et al. 2018).

### 1.2. Binary Choice Decisions

This prior research (Almaatouq et al. 2020, Atanasov et al. 2017, Becker et al. 2017, Farrell 2011, Gürçay et al. 2015, Jayles et al. 2017, Minson et al. 2018, Navajas et al. 2018) provides a compelling argument that social influence improves belief accuracy. However, as discussed above, numbers are not decisions, and numeric quantities are often converted to binary choice decisions by comparison to critical benchmark thresholds.

In our opening example, a manager needs to estimate the uncertain cost of building a software project and compare this quantity against the known cost of buying a pre-built product, and thus the 'buy' cost represents a critical benchmark threshold. Additional examples from related research further illustrate the diversity of decisions that are based on thresholds. Csaszar and Eggers (2013) study a theoretical model of decision-making in which managers must choose to accept or reject a project based on whether the project meets a sufficient quality threshold. More generically, threshold decision-making is illustrated by the behavior of "satisficing," a decision heuristic in which a person or organization adopts the first available solution that meets a minimum threshold of acceptability, as illustrated by a case study of an organization trying to find a new location for a warehouse (Cyert and March 1963).

While these examples focus on operational issues, additional examples show how threshold related forecasts inform strategic decisions. In financial markets, firms which report a "negative earnings surprise"—i.e., an earnings report below expectations—are subject to sudden shifts in perceived value even when falling short by as little as US$0.01 (Kinney et al. 2002), and thus these reported expectations represent an important threshold. Another example is US presidential elections. While it is common practice to forecast the number of electoral college votes a candidate may receive, the only practical outcome is determined by a threshold: whether a candidate receives more than 270 votes, the minimum (and sufficient) number of electoral votes needed to win.

Economic and political forecasting is also driven by psychological thresholds in addition to formal thresholds. For example, predicting whether or not a market or index will break a new record (i.e., pass the threshold marked by a previous high point) or even pass a round number may have an impact on investment decisions (Aggarwal and Lucey 2007, Schnusenberg 2006). For example, when the price of oil first passed the psychologically important threshold of US$100 per barrel it was reported as a major event by news media (Krauss 2008) and may have had an impact on market behavior more broadly (Parayitam and Dooley 2007).

## 2. The Crowd Classification Problem

When people are independent, binary choice decisions offer the same statistical advantages as numeric estimates, i.e. the wisdom of crowds. For example, the collected intelligence of many independent physicians can outperform the best individual physicians in generating cancer diagnoses when structured as a binary classification task (Kurvers et al. 2016, Wolf et al. 2015). Hastie and Kameda (2005) used simulation to study a hypothetical scenario in which groups must identify the highest-payoff option among several alternatives, finding that the averaging rule (choose the option with the highest average numeric estimate) performs roughly as well as the majority vote rule. Csaszar and Eggers (2013) examined this finding under more general assumptions, finding that voting can even outperform averaging in some circumstances.

The differences between numeric estimate and binary choice become more stark, however, when people are not independent. For numeric estimates, opinion sharing leads to convergence toward the average answer (Dalkey & Helmer 1963, Lorenz et al. 2011, Sherif 1935). As discussed above, this convergence can improve individual estimates with no impact on the group average or even improve the group average. In binary choice, however, the "average"—a vote tally or percentage—is not actually an opinion any person can hold. I.e. in the "build or buy" example, a group might vote 60% in favor of building, but a person cannot vote to build only a percentage of the software, nor can a group make their final decision as a percentage. They must simply choose either "build" or "buy." Thus while numeric estimates can converge to the average (preserving or improving the wisdom of crowds), convergence in binary voting leads to 100% agreement on one or another option. As a result, the social dynamics for numeric estimates and binary choice are starkly different despite showing similar properties for independent contributors.

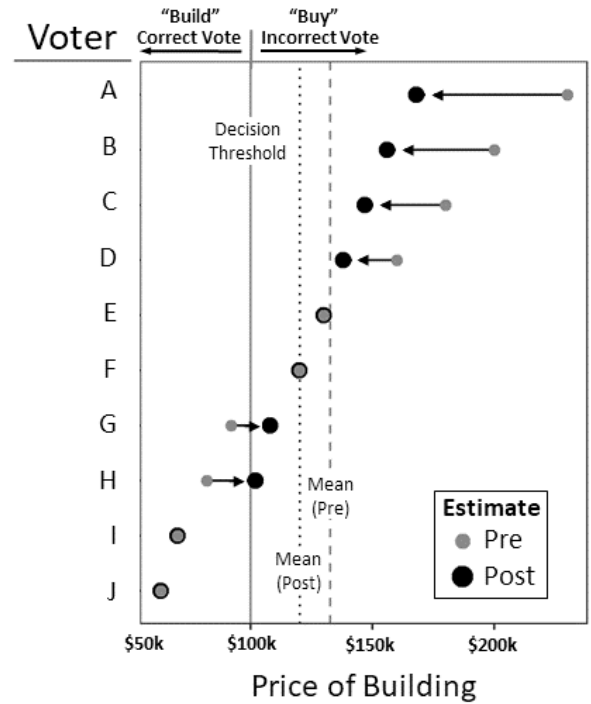

**Figure 1. Conceptual illustration of crowd classification problem for numeric estimates.**

*Notes. Points indicate beliefs pre- and post-discussion. Agents vote "buy" if their estimate is over 100, but a correct vote is "build." The dashed line shows the mean estimate pre- and post-discussion, getting closer to the true answer of "build" (less than $100k). However, the total number of people voting for "buy" increases from pre- to post-discussion.*

One intuitive strategy thus might be to encourage people to discuss numeric estimates before taking a vote on a binary choice. However, even this process—forming and exchanging numeric estimates prior to taking a vote—can grow the majority vote share, regardless of the accuracy of that vote. Moreover, the numeric estimate can become more accurate even as the binary vote becomes less accurate.

This seemingly paradoxical dynamic can be illustrated with a simple example, shown in Figure 1. Consider a team deciding whether to buy a client management database for $100k or build one in-house. Assume the true cost of building is less than $100k—they should buy (this truth is unknown to the team). Figure 1 shows an example of how the distribution of estimates might look before and after discussion. Grey points indicate pre-discussion estimates and black points represent post-discussion estimates. (Some members' pre-discussion and post-discussion estimates are the same, indicated by grey points on top of black points.) The only people who change votes are those whose estimates cross the threshold (voters *G* and *H*). Before discussion, a poll of numeric estimates ("how much will it cost to build?") shows that the mean estimate is $132.5k, meaning it's too expensive, and also shows that 60% of committee members

would vote against building. After discussion, the mean estimate is $120k—closer to the true value, and closer to the point of ambiguity—but now 80% of the committee members would vote against building.

Thus while the mean estimates might convey that the group after discussion has become less confident in buying, the size of the majority vote has increased, signaling increased confidence despite decreased voting accuracy. Below, after reviewing related binary choice models, we identify analytically the conditions under which this paradoxical outcome will occur and show that these conditions are consistent with commonly observed properties of empirical data.

## 3. Theoretical Analyses

This paper investigates whether the benefits of social exchange observed for continuous numeric estimation extend to binary choice estimation. We study two distinct processes. (1) In binary choice with binary communication ("binary exchange"), agents answer a binary choice question and observe the binary responses of other agents before providing a final vote. (2) In the exchange of numeric estimates prior to a vote ("numeric exchange"), agents first answer a continuous numeric question and observe the responses of other agents. Agents then update their numeric estimate and provide a final binary vote based on a threshold decision, i.e. whether their estimate is above or below some critical value.

In both models, our primary interest is the change in the vote. We will refer to a majority vote as growing (shrinking) if the number of people holding that belief increases (decreases). We note that this terminology is independent of accuracy, and our key finding is that the majority will grow in most circumstances. We will therefore additionally refer to the 'accuracy' of a binary vote, as measured as the percentage of people with the correct vote. We will thus refer to a group as getting more (less) accurate if the number of people holding the correct vote increases (decreases).

### 3.1. Binary Exchange

We study the following agent-based model.

1. Each agent $i \in \{1 \ldots N\}$ begins with a binary belief $B_i \in \{0,1\}$ distributed as binomial($N,P$).
2. In each timestep $t>0$, each agent $i$ flips belief (i.e. adopts belief $1 - B_i$) with probability $F_{maj}$ if their current belief $B_i$ is the majority opinion, and $F_{min}$ otherwise.
3. The majority opinion is defined as 1 if S>0.5 and 0 otherwise, where $S=\sum W_i B_i$, $\sum W_i=1$, and $W_i$ defines the amount of weight other agents place on agent $i$. (S is the influence-weighted vote.)

*Proposition 1a. When $W_i$ is equal for each agent (they are all equally influential) initial majority is more likely to grow than to shrink, and the initial majority opinion is an equilibrium opinion.*

*Proposition 1b. When $W_i$ varies, the unweighted majority opinion will be favored as long as the inequality $\Pi/(1-\Pi) > R$ is satisfied, where $\Pi$ is the proportion holding the initial majority and R is the ratio between the average weight given to minority-belief-holders and the average weight given to majority-belief-holders.*

The proofs follow algebraically from the assumptions and are given in the e-companion Appendix. When $F_{maj}=0$, it follows trivially that the majority opinion will grow until all agents have adopted the initial majority in an absorbing state of 100% agreement, or consensus, since agents in the majority will never change their opinion. While it may not be common that people will change their opinion if the majority already agrees with them, for thoroughness we also consider the case where $F_{maj}>0$. In this case the group will not reach an absorbing state of consensus, because there will always be some chance that an agent will flip, but as long as $F_{min}>F_{maj}$ the group will reach an equilibrium that maintains the initial majority. We note that these results hold in expectation, meaning that small groups may of course deviate due to sample variation. Proposition 1 is therefore presented to characterize some basic dynamics of group behavior and is not intended as a deterministic law of group behavior.

**3.1.1 Discussion.** Empirical research identified variation in response to social information (e.g., stubbornness) as an important mechanism in determining the effect of social exchange on group accuracy. However, a confident minority cannot overturn a less confident majority from this effect alone (see e-companion Appendix). Thus binary choice dynamics will favor an initial majority opinion regardless of accuracy, even under conditions that would improve the accuracy of numeric estimates. This claim is not itself surprising and is consistent with simpler models of binary choice such as the voter model (Mossel and Tamuz 2017). However, when considered in the context of belief accuracy, these majority-rules dynamics impose substantial limitations on prior claims (Becker et al. 2017, Farrell 2011, Jayles et al. 2017) that social information exchange improves accuracy for both groups and individuals. At the same time, however, the scope conditions identified in Proposition 1b show a possible solution to the crowd classification problem. While the empirical analysis in the present paper is designed to demonstrate the existence of the crowd classification problem, we discuss in the conclusion how these scope conditions offer guidelines for future empirical research.

**3.2. Numeric Exchange**

**3.2.1. Summary.** In this analysis, we formally describe the conditions generating the proof-of-concept example shown in Figure 1. This analysis consists of three parts. Proposition 2a describes the general conditions under which the initial majority will either grow or shrink. As in the preceding model, a majority is said to 'grow' if the number of people holding the majority opinion increases; and the majority 'shrinks' if the number of people holding the majority opinion decreases. Critically, these conditions can be described without reference to the true answer. This characteristic is the heart of the crowd classification problem: binary choice belief dynamics are driven by the initial distribution, and not the true answer.

Proposition 2b illustrates a general set of conditions under which the mean numeric estimate becomes more accurate even as the binary vote becomes less accurate. As above, we describe a vote as becoming less (more) accurate if the number of people holding the correct answer decreases (increases). The conditions given in Proposition 2b are consistent with common empirical conditions and serves as an example of how the crowd classification problem may reverse the expected benefits of social influence. Notably, the conditions described in Proposition 2b are not the only way in which a vote may become less accurate, and are intended to serve only as an illustrative demonstration of the social dynamics of belief formation. Excepting the general result presented in Proposition 2a, we do not provide an exhaustive demonstration of all possible parameter combinations. The framework presented here can be readily adapted to address any particular scenario of interest.

Finally, Proposition 2c extends the asymptotic results of Proposition 2a and 2b to finite-time outcomes. Whereas Proposition 2a and 2b assumes that groups converge on a single shared consensus estimate, Proposition 2c shows that similar results are obtained when using the post-discussion mean as a proxy for the asymptotic consensus estimate. In order to test the validity of assumptions made to support Proposition 2c, we further test these outcomes with numerical simulation.

**3.2.2. Model.** We now consider the dynamics of the DeGroot (1974) model for N agents, with the added assumption that agents at the end convert their numeric estimate to a binary vote according to a threshold rule. This model can be described as follows:

1. Each agent $i \in \{1 \ldots N\}$ begins with a numeric belief $B_{i,t=0} \in \mathbb{R}$.
2. Each agent *i* places some fixed weight $W_{i,j}$ on the belief of agent $j \in \{1 \ldots N\}$ (including weight $W_{i,i}$ on their own belief) such that $\sum_{j=1}^{N} W_{i,j} = 1$.
3. At each timestep *t+1*, each agent's new belief is a weighted sum across all other agent's beliefs, i.e. $B_{i,t+1} = \sum_{j=1}^{N} W_{i,j} * B_{j,t}$.
4. Whenever the process is stopped (we discuss both asymptotic and short-term outcomes) each agent will vote according to a threshold T, such that their vote is determined by whether $B_{i,t} > T$.

*W* represents a matrix of weights that can be treated as a network adjacency matrix indicating how much weight each agent places on each other agent. If agent *i* either does not observe or ignores agent *j*, then $W_{i,j}=0$.

**3.2.3. Basic Dynamics.** DeGroot (1974) demonstrated that if this network is aperiodic and consists of a single component, the group will asymptotically converge so that every agent holds the same belief, which we here term the consensus belief. For comparing pre- and post-discussion accuracy, we focus on the mean belief as the 'collective belief' (Page 2007) because it is the primary outcome of interest to many researchers studying the wisdom of crowds in networks (Almaatouq, Noriega-Campero, et al. 2020, Atanasov et al. 2017, Becker et al. 2017, Da and Huang 2020, Frey and van de Rijt 2020, Lorenz et al. 2011).

We show that the dynamics of the binary vote do not depend on the relative location of the initial and final mean belief. Instead, we find that the dynamics of the binary vote depend only on the relative location of the median, consensus belief, and decision threshold. This includes the case where the mean belief does not move, i.e. where social influence has no effect at all on the average belief. For example, in a facilitated process where everyone is equally influential ($W_{i,j} = constant$ for all i,j) the consensus belief will be the same as the initial mean. Critically, however, as individuals converge toward this pre-discussion mean their individual votes will change. This dynamic results in the crowd classification problem.

Formally we can state this claim as follows. Let M indicate the pre-discussion median belief; let C indicate the consensus belief; and let T indicate the threshold on which the binary vote is based, then:

*Proposition 2a. In continuous numeric information exchange reaching asymptotic consensus, the majority vote according to a threshold decision will grow except when C<T<M or M<T<C—i.e., the majority vote will shrink if and only if the threshold falls between the initial median and the consensus.*

The proof follows from the assumptions. First observe that the median estimate indicates where the majority opinion will fall relative to the threshold—i.e., what the majority vote will be. Prior to discussion this is given by M, and after discussion this is given by C. Therefore, when C<T<M or when M<T<C, the majority opinion will switch sides relative to T, by assumption, as the majority vote moves from M to C. In contrast when M<C<T, C<M<T, T<M<C, or T<C<M (i.e., when M and C are on the same side of T) then the majority opinion will be on the same side of T both before and after discussion. And as consensus by definition yields 100% agreement on votes, a majority opinion that is the same at the end has grown.

**3.2.4. Relationship to accuracy.** Becker, Brackbill and Centola (2017) showed empirically and in numeric simulations that if individual agents' self-weight ($W_{i,i}$), i.e. stubbornness, is positively correlated with accuracy—such that people with greater accuracy are more stubborn and therefore make smaller revisions—then the mean belief in a group will become more accurate over time as a result of information exchange. One implication of Proposition 2a is that the dynamics of the majority vote are decoupled from truth, since they are determined only by M, C, and T. As a result, it is possible for the mean belief to become more accurate even as the vote itself becomes less accurate. This scenario will occur when the initial majority vote is inaccurate, the conditions of proposition 2a are met (so the majority grows), and the mean becomes more accurate.

We can construct a formal possibility proof of this scenario by examining one possible set of conditions leading to this outcome. Let $\mu$ be the pre-discussion mean belief, and let $\theta$ be the true answer, then:

*Proposition 2b. The mean numeric belief will become more accurate even as the binary choice vote becomes less accurate when* $M<\mu<C<T<\theta$ *or the symmetrical case* $\theta<T<C<\mu<M$

Proof: Consider the case $M<\mu<C<T<\theta$, with the same argument applying to the symmetrical case. First, observe that the majority vote is initially inaccurate whenever $M<T<\theta$ or $\theta<T<M$—i.e., when the threshold falls between the initial median and the true value. This is true because when the threshold falls between M and $\theta$, the majority vote (determined by M) is on the opposite side of the threshold from $\theta$, i.e. the majority gives an incorrect vote. Therefore if it is true that $M<\mu<C<T<\theta$ then it will be true that the majority belief is initially inaccurate. And, by Proposition 1a, the majority vote will become amplified (more inaccurate i.e. less accurate). Moreover, it's true both that $\mu<C<\theta$ and that $M<C<\theta$—i.e., the consensus belief (equal to the post-discussion mean and median) is closer to the true belief than either the pre-discussion mean or the pre-discussion median. In other words, both the mean and median numeric belief become more accurate as the vote becomes less accurate. As we discuss below, these conditions are both theoretically illustrative and also consistent with empirical data.

**3.2.5 Short Term Dynamics.** The chief limitation is that this proof considers only asymptotic (consensus) outcomes and does not describe the case where people still disagree at the end. We next consider short-term outcomes. In the short term, there is no 'consensus.' In this case, let C represent the mean belief after some number of revisions. We first make the simplifying assumption that any individual's belief change over time is monotonic in the direction of C. Then, it is trivial to argue that the conditions described in proposition 2b hold for short term dynamics. We verify the validity of this approximation with numeric simulations presented in the e-companion Appendix.

*Proposition 2c. Even for short-term dynamics, the conditions described in proposition 2a and 2b hold.*

Proof: We first make the simplifying assumption that individual belief change over time is monotonic. Note that the only people whose beliefs will change the final vote count are those individuals whose numeric beliefs cross the critical decision threshold. And by the assumption of monotonicity, the only people who change their votes will be those people who have an initial vote other than C, i.e. an initial vote different from what *would* be the consensus vote in an asymptotic model. Thus, the only people who flip votes are those who initially disagree with C. Therefore, the number of people supporting C only grows and the number of people disagreeing with C only shrinks—thus, short term dynamics produce the same outcome as consensus dynamics.

While monotonicity is an empirically plausible behavioral assumption, there are theoretical edge-cases in the DeGroot model in which an individual's belief trajectory is non-monotonic with respect to time. To test the fitness of our approximation, we conducted 10,000 numeric simulations (i.e. Monte Carlo experiments) of the DeGroot model with N=1000 agents and 10 revisions with normally distributed beliefs and uniformly distributed self-weight $W_{i,i}$. Our simulations results, reported in the e-companion Appendix, generally support the simplifying assumption and resulting argument presented in Proposition 1c (see Fig. A1 in Appendix section EC.4.).

Across most of the range of thresholds (0 to 1) our model fits 100% of simulated experiments. However, when the threshold is very close to the initial mean, the approximation breaks down slightly (see Fig. A1 in e-companion Appendix) with about 2.5% of simulated experiments deviating from predictions. This deviation can be explained by the observation that as the mean belief changes, it will cross the points of some initial individual beliefs. Those individuals are the ones who will fail the monotonicity assumption in theory (if not empirically) as they move first one direction then another to follow the mean belief. Finally, note that the individuals nearest the threshold are most likely to cross the threshold and thus change the vote. Thus when the threshold is near the mean, those individuals who fail the monotonicity assumption are also the individuals who change their votes and thus determine the change in majority.

**3.2.6 Discussion.** Proposition 2b identifies 'worst case' conditions under which the mean belief gets more accurate even as the binary vote gets less accurate. The conditions described in Proposition 2b are not the only conditions that can lead the vote to become less accurate, and Proposition 2b is not intended as an exhaustive description of the social dynamics of binary estimates. This particular example was chosen because it is broadly consistent with empirical data, as described in our empirical analysis below.

The crux of the crowd classification problem is captured by Proposition 2a. At the heart of the crowd classification problem is the observation that the dynamics of the binary vote are decoupled from the dynamics of the numeric estimate. Notably, we identify two broad areas of the parameter space that determine whether the initial majority grows or shrinks. Importantly, when M=C—e.g., for a normal belief distribution where the group converges on the initial average—the majority will always be amplified. Empirically, numeric beliefs often follow a skewed distribution, opening the possibility for the majority vote to shrink or even flip. Still, these dynamics also represent the crowd classification problem, as the crowd vote remains decoupled from the crowd numeric estimate.

## 4. Empirical Analysis: Binary Exchange

### 4.1. Methods

To test Proposition 1, we conducted a pre-registered experiment. This experiment follows a procedure similar to prior research on belief accuracy (Almaatouq, Noriega-Campero, et al. 2020, Becker et al. 2017, Lorenz et al. 2011) in which subjects answer factual questions before and after being able to observe the responses of other subjects. Our estimation tasks consisted of a numeric estimate with a threshold-based binary choice. For example, one question asked subjects to estimate the number of Americans who think that science and technology improves our lives. Subjects were given two response options, "Above 60%" or "Below 60%". Subjects were paid based on the accuracy of their answers. A single trial consisted of 20 subjects simultaneously responding to a single question. We collected data for 5 unique tasks with 3 different thresholds each. We replicated each unique task 4 times, collecting data on 60 trials total. Detailed methods, question wording, and screenshots are provided in the e-companion Appendix.

Prior to conducting our experiment, we conducted a pilot test designed to estimate individual behavior to support statistical power tests and generate our pre-registered predictions. The methods and results are described in the e-companion Appendix. The primary goal of this pilot study was to estimate the function $P(x)$, or the probability that a subject on our platform will change their answer when $x$% of people disagree with them. The estimates for individual behavior were used to simulate an empirically calibrated agent-based model. The results of this simulation were pre-registered and used to generate Figure 2, allowing us to compare our experimental results with theoretical predictions.

### 4.2. Results

Our experimental results support our theoretical predictions. Figure 2 shows our experimental results (colored points, red line) compared against our theoretical predictions (black line). On average, groups with an initially inaccurate vote (N=26) decreased in accuracy by 6.3 percentage points ($P<0.02$, Wilcoxon rank sum test), while groups that were initially accurate (N=33) increased in accuracy by 9.3 percentage points ($P<0.001$, Wilcoxon rank sum test) and the two conditions were significantly different ($P<0.001$, Wilcoxon rank sum test). While our pre-registered analysis erroneously included the single 50/50 split group as initially accurate for this two-sample comparison, we omit that datapoint. Results are comparable if that datapoint is included, and both analyses are included in the supplemental materials.

These tests are relatively conservative, since they include outcomes for groups in which the average belief (proportion voting for each answer) did not change at all. Table 1 shows the total count of each outcome (increase in accuracy, decrease in accuracy, or no change) conditional on initial accuracy (accurate, inaccurate, or split). Conditional on showing any change at all in the majority, 81% of the initially inaccurate groups became even less accurate ($P<0.01$, proportion test), with an average accuracy decrease

CROWD CLASSIFICATION PROBLEM                                                                                                    16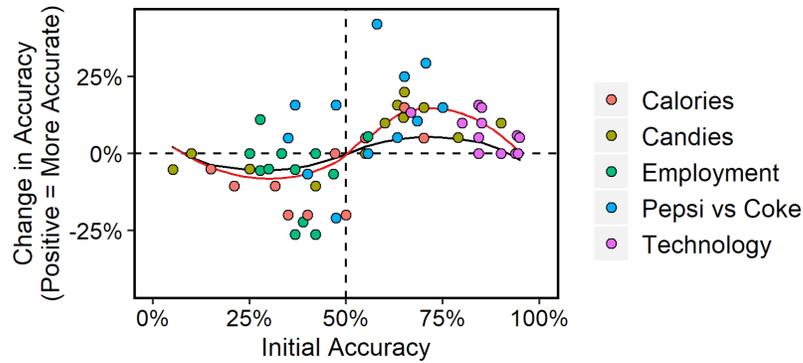

**Figure 2. Change in Group Accuracy as a Function of Initial Group Accuracy**

*Notes: Black line indicates simulation results, colored points indicate experimental results, and red line indicates a polynomial best fit to the experimental data.*

|                      | Initial Accuracy |            |       |
|---------------------:|:----------------:|:----------:|:-----:|
|                      | Accurate         | Inaccurate | Split |
| Increased Accuracy   | 27               | 4          | 0     |
| Decreased Accuracy   | 0                | 17         | 1     |
| Unchanged            | 6                | 5          | 0     |

**Table 1. Count of Group Votes by Initial Accuracy and Change in Accuracy**

of 7.8 percentage points. In contrast, 100% of the initially accurate groups, conditional upon showing any change at all, became more accurate ($P<0.001$, proportion test) with an average increased accuracy of 12.4 percentage points. We provide a more detailed analysis of the model fit in the e-companion Appendix.

**4.2.1 Relationship between accuracy and revision.** In research on the social dynamics of numeric estimates, one important factor is the correlation between response to social influence (or confidence) and accuracy. In our pilot data, we observe an effect of accuracy, such that people were less likely to revise their answer if they were initially accurate ($P<0.09$, logistic regression controlling for observed disagreement). Overall, 19% of individuals who were initially inaccurate revised their answer, while only 13% of those individuals who were initially accurate revised their answer. Results are comparable but statistically stronger in our main experiment: 21% of initially inaccurate individuals revised their answer after observing social information, while only 13% of initially accurate individuals revised ($P<0.001$, proportion test). In numeric estimation, such a correlation is sufficient to improve the accuracy of the group average (Becker et al. 2017, Madirolas and de Polavieja 2015). As expected from our theoretical analysis, however, this correlation is not sufficient to improve the accuracy of the group vote.

### 4.3. Discussion

This experiment demonstrates the main claim of this paper: the benefits of collective intelligence via social information processing, previously observed in several studies on numeric estimates (Almaatouq et al. 2020, Atanasov et al. 2017, Becker et al. 2017, Farrell 2011, Gürçay et al. 2015, Jayles et al. 2017, Minson et al. 2018, Navajas et al. 2018), do not extend to binary choice estimates without further intervention. One aspect of majority amplification highlights a unique risk of binary choice estimates in small groups such as boards, task forces, and committees. If initial beliefs are close to evenly split, a committee vote may favor one or another decision simply by chance variation. While this initial random chance may tip the scales only slightly, social influence can turn this minor majority into an apparently robust democratic mandate. In one experimental trial, we observed an initial accuracy of 47%. Had we polled those people on another day, or had one person failed to offer an answer, then the vote may have turned out differently. Nonetheless, this slight 53% majority (in favor of the incorrect opinion) ballooned to a robust 74% support after group interaction. This increased agreement could encourage additional (unfounded) confidence in the group's inaccurate collective decision.

## 5. Empirical Analysis: Numeric Exchange

### 5.1. Methods

To test Proposition 2, we analyze data made publicly available as supplementary materials with research published by Gürçay, Mellers, & Baron (2015), Becker, Brackbill, and Centola (2017), and Lorenz et al. (2011). The experiments by Becker et al. and Lorenz et al. followed a nearly identical procedure to our experiment on binary exchange but with numeric estimation tasks. Full details on their method can be found in their publications. Gürçay et al. followed similar methods, but in addition to sharing numerical estimates they also provided subjects with text chat interfaces (i.e., allowing free discussion). Gürçay et al. also asked subjects to report confidence, and subjects were able to see each other's reported confidence alongside the numeric information chat text. In brief, subjects in both studies provided independent estimates, then were exposed to information about the beliefs of other subjects in the trial, and then provided final estimates.

To measure whether empirical data is consistent with Proposition 2, we first classify a trial based on the conditions that predict whether or not the initial majority vote share will grow. These conditions are given by Proposition 2a, applied to short-term dynamics in Proposition 2c: the majority vote will shrink only when the threshold (T) falls between the initial median (M) and the final mean (C), i.e. when $C<T<M$ or $M<T<C$. Otherwise, the majority vote will be amplified. After determining whether a trial meets these conditions—i.e., whether the majority vote is expected to grow—we then determine whether or not the majority vote actually grew in the empirical data. That is, we use this condition to determine whether empirical data behaves as theoretically expected.

We define a trial as a single group answering a single question before and after social influence. Thus the median (M) and the mean (C) are given values for each trial. However, the decision-threshold T is not given by the data itself and must be assumed. We therefore test data for a range of possible thresholds for each trial. For a given threshold, we ask the hypothetical question: what would subjects have voted according to their numeric estimates before and after social influence? Thus for any particular assumed value of T we can ask for any a trial (given M and C) whether the majority vote share was predicted to grow and also whether the majority vote share actually grew.

**5.1.1. Analysis.** One challenge is that a given threshold is not comparable across different questions. For example, very few people will think that the average temperature in January in London is over 100, but that might be a reasonable guess for the count of candies in a jar. Thus to normalize thresholds across questions, we measure thresholds not in the original units of the question but in quantiles. Suppose, hypothetically, that the $80^{th}$ percentile response for temperature is 30 but the $80^{th}$ percentile response for candies is 200. We would consider those two thresholds to be equivalent, since they reflect equivalent percentiles despite being different numeric values in the original units.

We can express this process formally using the quantile function. The quantile function Q(P) indicates the value for any numeric distribution such that P% of people hold a belief under that value. For the example above, Q(80%)=30 for the temperature question and Q(80%)=200 for the candies question. Thus for any given trial and any given value P from the range 0% to 100%, we can set T=Q(P). In this context, the quantile function asks: what is the decision threshold T=Q(P) such that P% of people provide an answer lower than T? Using this threshold value T, we can then measure the accuracy of people's hypothetical votes before and after social influence.

Thus given values M and C provided by the data, and an arbitrary T=Q(P), we can measure our two key outcomes: (1) whether this trial/threshold pair meets the conditions given by Proposition 2, and (2) whether the majority as measured by T increased or decreased. Finally, we set T=Q(P) for the full range of values 0% to 100%, and test whether Proposition 2 correctly characterizes outcomes for each possible threshold. This process allows us to compare outcomes across trials with different questions by standardizing thresholds according to the percentile rather than the numeric value.

To help interpret these results, we take advantage of the relationship between the quantile function P and group accuracy. Specifically, we note that each value P corresponds to an initial accuracy of either P or 100-P. This can be seen by observing that the threshold divides people's initial numeric estimates into two camps—accurate and inaccurate—and the quantile function thus measures the percent on the accurate/inaccurate sides of the threshold. Thus to aid interpretability, we visually organize data according to initial accuracy—i.e., treating initial accuracy as the independent variable—since these rates correspond to the quantile values P.

For each experimental trial in our data reanalysis, we empirically assess the numeric threshold $T=Q(P)$ based on pre-discussion responses for all values P from 1% to 99% in increments of 1%. For each value of $T=Q(P)$, we measure the accuracy both before and after social influence. We then aggregate across all trials (for a given quantile threshold value P) and calculate the average change in accuracy as a function of initial accuracy. Critically, however, we divide outcomes (red/black) by the conditions given in Proposition 2—whether the majority is expected to grow (black) or shrink (red). This process allows us to represent the theoretical framework given by Proposition 2 in the same way as we represent outcomes for experimental data on binary exchange.

## 5.2. Results

**5.2.1. Binary Vote Dynamics.** Figure 3 shows the change in accuracy as a result of initial accuracy for data in which the majority is predicted to grow (black lines) and data in which the majority is expected to shrink (red lines). This figure shows that the majority grew for trial/threshold combinations which meet the conditions described in Proposition 2a (black lines): those trials which were initially accurate became more accurate, and those trials which were initially inaccurate became less accurate. In contrast, trials which did not meet the conditions in Proposition 2a (red lines) show the reverse. As can be seen visually in Figure 3, Proposition 2 is less accurately predictive when the threshold is near the median, consistent with our discussion on numeric simulations of short-term outcomes (see also Figure A1, e-companion Appendix).

**5.2.2. Theoretical Fit.** In order to quantify the model fit, we calculate—for each independent trial—the percentage of threshold values for which our theoretical model correctly predicted outcomes. That is, for the 99 values of P ranging from 1% to 99% we count how many empirical outcomes matched theoretical predictions. We note that results are robust to variations in the 'resolution' of the threshold values (see Table A1, e-companion Appendix). This process produces a single measure of fit (percentage of trial-threshold combinations correctly predicted) for each independent experimental trial. Across all trials, an average of 79% (2%), 76% (1%), and 84% (2%) of outcomes (standard error in parentheses) were accurately predicted for the Lorenz et al., Gurcay et al., and Becker et al. datasets. Note that our goal is not to produce the best-fitting model, but to test whether empirical data is broadly consistent with our theoretical argument. These results suggest that the theoretical prediction in Proposition 2 is broadly consistent with the qualitative dynamics shown in each empirical dataset.

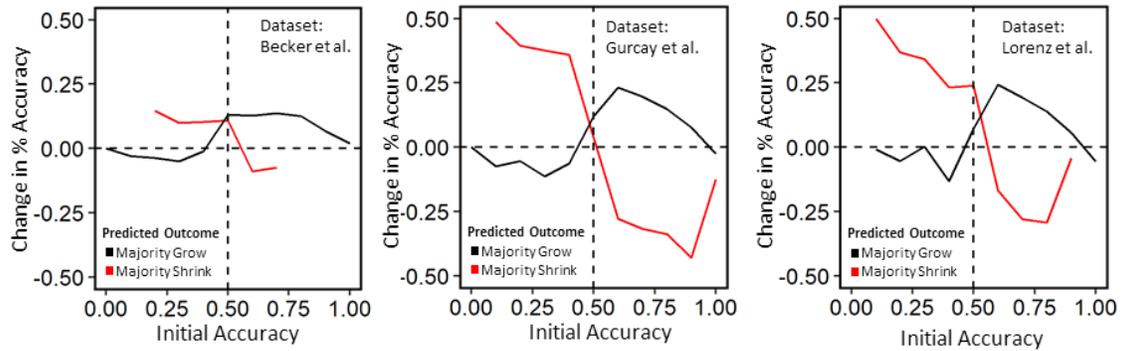

**Figure 3. Empirical analysis testing Proposition 2 for two previously published datasets.**

*Notes: For each trial, we test thresholds from 0 to 1 standardized across estimation tasks with the quantile function Q(P). Each threshold P corresponds to an accuracy of either P or (1-P) as described in the main text. This figure shows the change in group accuracy as a function of their initial accuracy, averaged across all trials for a given threshold value P. The black (red) line shows outcomes for trial/threshold pairs for which Proposition 2 predicts the majority will grow (shrink).*

One possible concern is that the crowd classification problem characterizes convergence dynamics in general, but that this problem is less descriptive when social learning takes place. That is, perhaps our predictions fail when groups become more accurate. Critically, we find no difference between those cases where the mean improves and those cases where the mean does not improve (Fig. A4, e-companion Appendix). This invariance to changes in numeric estimate accuracy supports our interpretation of the crowd classification problem as a de-coupling between the binary vote and the numeric estimate and a tendency for the majority to grow regardless of accuracy.

**5.2.3. Assessing the Problem.** To assess the extent to which the crowd classification problem is likely to pose an issue, we measure the proportion of trials that fall into the different categories identified by Propositions 2a and 2b. The primary condition of interest is given in proposition 2a, which states that the majority will grow *except* when the threshold falls between the initial median and the final mean. While we cannot offer a theoretical basis for determining where the threshold is likely to fall in practical scenarios, we can measure the proportion of total answers which fall in this range. Across all trials in all datasets, we find that 21% of individuals provide a response that falls between their group's initial median and final mean. Thus if a decision threshold were selected at random from among individual estimates (an illustrative hypothetical only) then groups on average would have a 79% chance of meeting the conditions that lead to majority growth.

The second scenario we can consider is Proposition 2b, in which the initial mean becomes more accurate. We find that the mean became more accurate in 72%, 55%, and 68% of trials collected by Lorenz et al., Gurcay et al., and Becker et al., respectively. This observation shows that, as expected from previous reports, social influence generally improves the accuracy of the mean estimate and that this outcome commonly occurs in the datasets we analyze. More specifically, the example given in Proposition 2b assumes that either $M<C<\theta$ or that $\theta<C<M$. We find that this condition is met over 50% of trials averaged across all datasets. These measurements show that empirical data is commonly exposed to the worst case scenario for majority-amplification dynamics as predicted by the crowd classification problem. We next test whether these conditions correspond to the observed binary voting dynamics.

### 5.3. Discussion

These findings support the general argument that social influence will tend to grow the initial majority vote share, regardless of accuracy. As a result, social influence will often (but not always) undermine group accuracy precisely when they most need the benefits of collective intelligence—when they are initially inaccurate. More broadly, these dynamics suggest that it is both possible and likely that—for a binary choice task with an underlying numeric estimate—a group vote will become less accurate even as the mean numeric belief becomes more accurate.

As one illustrative example, consider a trial in our re-analysis of the data by Becker, Brackbill, and Centola (2017) in which individuals were estimating the number of calories in a meal. If you had asked the group whether or not there were fewer than 600 calories in the meal (truth=729 calories) then 74% of subjects would have initially said "yes" (incorrect) while the mean estimate was 435 calories. After observing each other's beliefs, the percentage voting "yes" increased to 84%—a stronger display of agreement around an incorrect answer—even as the mean belief shifted to 466 calories, closer to the decision threshold. Thus while the majority vote sends a more confident signal after social influence, the mean belief has moved closer to the decision threshold, indicating that an observer should doubt the group belief and seek additional information.

### 6. General Discussion

Our primary contribution is the demonstration that social exchange processes which reliably increase numeric estimate accuracy will not generally increase the accuracy of decisions based on that estimate. Critically we find that social influence can lead a group's binary vote to increasingly favor the wrong choice even as the average numeric estimate becomes more accurate. The real challenge, however, is that groups do not know in advance whether social influence will help or harm voting accuracy.

At the heart of the crowd classification problem is a broad tendency to grow the initial majority. In binary exchange, we argue theoretically and find empirically that an initial majority is nearly always amplified. For numeric exchange, we find that symmetrical dynamics may sometimes occur in which the majority shrinks, but that empirical belief distributions suggest that majority-rules will be the most common outcome. Despite the qualitative process differences between binary exchange and numeric exchange, the similarities between Figure 2 and Figure 3 highlight this shared pattern: both show an 'S' curve relating initial accuracy to change in accuracy. Although these curves are generated by starkly different underlying dynamics, they nonetheless reveal a common outcome pattern for both types of group behavior.

## 6.1. Practical Implications

Our research is most informative for groups who are actively trying to optimize their decision-making, noting that decisions in organizations occur through both planned and haphazard processes (March 1991). When we discuss process optimization, we assume that a group has already decided to approach their decision with some intention. One central assumption is that the question itself is clearly defined, and that the group has reached the point of identifying a focally important numeric estimate. We therefore note that we provide a prescriptive approach to decision-making, rather than a descriptive approach.

Consider again the opening example of a manager faced with the "build or buy" software development decision. In this example, the manager must compare an uncertain quantity, the cost of building a custom solution, against a known quantity, the cost of purchasing a commercial solution. A simple strategy to improve estimation accuracy is to harness the wisdom of crowds—i.e., poll multiple team members on their opinion and use the average answer. If this manager wishes to further improve their accuracy, they may encourage their team members to first produce an independent estimate, then discuss their opinions, and then provide a final post-discussion estimate. Whereas previous research would have offered this advice as a blanket strategy, our research offers both an important cautionary caveat and a simple corrective: do not poll team members directly on their binary "build or buy" opinion, and actively discourage them from discussing this binary opinion with each other. Instead, focus team discussion on the cost value (numeric estimate) itself.

Our research has more critical implications for those groups for whom voting is an integral part of their process. Previous research suggests both that voting and averaging yield similar results when opinions are aggregated independently (Csaszar and Eggers 2013, Hastie and Kameda 2005) and that social influence improves opinion accuracy (Almaatouq, Noriega-Campero, et al. 2020, Becker et al. 2017, Farrell 2011). Critically, however, we find that voting and averaging produce dramatically divergent interactions with social influence: social influence improves numeric accuracy but undermines vote accuracy. Thus when voting is required by norms or governance, our research suggests that such votes should be contributed independently in order to maximize the wisdom of crowds.

Based on prior research, a manager observing increased agreement after discussion might conclude that they can proceed with confidence in their group's decision, but the crowd classification problem reveals that this confidence would be misleading. By preserving the independence of votes, rather than reaching consensus, a group can more accurately reflect its relative certainty (or lack thereof). In those cases where social independence is not possible—e.g. where discussion or interaction is unavoidable—one strategy is to avoid discussing any opinions or judgements (either numeric estimates or votes). Instead, groups could focus discussion instead on sharing facts or other information without evaluation, such that people form opinions independently based on the available information.

While our analysis focuses on threshold-based decisions based on single numeric estimate, other forecasting or estimation-based decisions may involve multiple uncertain values. As we discussed in our opening example of market entry, prescriptive decision theory (e.g. negotiation theory) often recommends strategies for reducing even complex decisions to single test metrics, as when comparing the expected payoff of a complex business deal against the best "no deal" alternative (Bazerman and Moore 1994, Medvec and Galinsky 2005). However, there may be additional complexities associated with multivariate estimation tasks that we do not consider here. Recent research has begun to demonstrate how crowds can collaboratively generate "system models" that specifically harnesses the complex structure of problems (Amipour et al., 2020). However, it's important to note that if groups are in a context where they can explore such creative decision processes, then they presumably also have the option to avoid the crowd classification problem by avoiding voting.

### 6.2. Complex Problem Solving and Collective Intelligence

Our research contributes to a growing body of theory and experiments studying "collective intelligence," which generally refers to the benefits that emerge from groups as compared with individuals. We note that the definition of collective intelligence varies across scholars, including the very broad definition "groups of individuals acting collectively in ways that seem intelligent" (Malone and Bernstein 2015). Nonetheless, the interdisciplinary body of researchers using this term all tend to share one thing in common: an interest in optimizing group processes. Our research aims to identify how to optimize group accuracy.

Even to the extent that our work can apply to more complex issues of accuracy such as multi-variate estimation, we nonetheless take as assumption that the options being evaluated—what to build, what to buy—have already been determined, and we assume that the only task is evaluating given solutions. We also assume that the payoff function is given in advance, i.e. that there is no conflict over which features matter to evaluation and what they're worth. In other words, we study the specific process of estimation in isolation. At the same time, other models and experiments have been developed to study other components of problem solving and decision-making in isolation. For example, research on innovation (Lazer and Friedman 2007, Shore et al. 2015) and brainstorming (Stroebe et al. 2010) examines how factors such as information sharing impact the number and quality of solutions a group can generate. However, just as we take for granted that solutions have been identified—and need only to be evaluated—these models often take for granted that the payoff for a found solution is known with certainty, and assume away the uncertainty that forms the central interest in our present research.

One strategy to integrate our research on estimation accuracy with other findings about group decision-making is to adopt a normative, prescriptive process that intentionally isolates components. For example, a group would first engage in an agenda setting or sensemaking process that defines the question or need; would then identify possible solutions; and finally would evaluate them, i.e. an estimation process. For each step, the group would follow best practice according to theory and evidence on that particular process. This strategy would be consistent with the theoretical perspective—supported by the present results, as well as related research (Almaatouq, Yin, et al. 2020, Becker et al. 2020, Lazer and Friedman 2007, Straub et al. 2020)—that different tasks are optimized by different processes. This strategy is also consistent with the methods of decision facilitation practitioners, who often isolate the steps of defining the problem, generating solutions, and evaluating/selecting solutions (Fisher et al. 2011, Fisher and Charkoudian 2008).

In practice, groups often follow more chaotic and emergent processes of decision-making (Cohen et al. 1972, Coleman 1966, Cyert and March 1963), and a full collective intelligence theory requires understanding how different types of tasks interact. Thus one important aim for future research will be to develop integrative models that explore how each of these tasks interact. For example, it may be possible to develop a model of problem-solving or brainstorming which also incorporates elements of estimation, in which a group must simultaneously or iteratively generate and evaluate ideas.

### 6.3. Additional Limitations

We note that neither our model nor our experiment addresses multiple choice scenarios. The many possible multiple choice voting procedures developed by democratic theorists (e.g. approval voting, ranking) introduce a large number of possible processes by which a group may reach a decision, and more complex models are needed to describe their social dynamics with respect to accuracy. Importantly, our findings show that we cannot generalize prior results on social learning—including the present findings—to such multiple choice voting systems. Rather, we show that the benefits and risks of social information processing depend highly on the structure of the task at hand. Notably, our findings provide a warning not to assume that increases in numeric accuracy translate to increases in decision accuracy. Our research highlights the importance of specifically examining the effect of social processes not only on binary choice but voting accuracy more broadly.

In addition to the limited scope regarding the type of problem we study, the experimental results described here (both new and re-analysis) reflect only limited forms of communication. In particular, we study the role of strictly informational influence. In the case of binary exchange, this influence is further restricted to opinion sharing and excludes confidence sharing and free discussion. However, beliefs and social influence in many empirical contexts are shaped not only by strictly informational influence but also by factors such as informational framing (Gardikiotis et al. 2005) and emotional contagion (Collins 2014). In an effort to identify the basic dynamics of group belief formation, our analysis necessarily abstracted away some mechanisms that could influence the variables or assumptions defined in our theoretical models.

We identify two key factors of interest that our current approach overlooks. First, we note that our model largely assumes decisions are made on some underlying numeric quantity. While prescriptive decision theory recommends that people quantify even subjective decisions, it would be valuable to understand how social influence impacts belief accuracy even in the absence of explicit quantification. In this respect, our first model and experiment offer some insight—suggesting that the initial majority would generally dominate. However, a second important factor is that people in practice may share detailed arguments which are inherently convincing, thus introducing additional avenues for influence. Convincing rhetoric, evidence, or argumentation is one possible way that an initial minority may overturn the initial majority. Nonetheless, we expect that such dynamics would be the exception and not the rule. For example, analyses of jury deliberations—which are unlikely to be explicitly quantified and which involve complex argumentation—suggest that majority-rules dynamics are the most likely determinant of outcomes (Burghardt et al. 2019).

**6.4. Towards a Solution to the Crowd Classification Problem**

One benefit of using such a simplified model is that our theoretical analysis precisely defines the scope conditions of our results, suggesting possible "solutions" to the crowd classification problem—i.e., the identification of conditions that might reliably improve vote accuracy.

For binary exchange, the scope conditions on Proposition 1 suggest that an accurate minority might overturn an inaccurate majority if sufficient weight is placed on accurate individuals. One design suggested by Proposition 1b would be a mechanism that lets people self-report confidence—or any other kind of mechanism that causes accurate individuals to also be persuasive. However, despite the positive correlation observed here, that correlation is also frequently negative for binary choice (Koriat, 2012). Thus future research may focus first on aligning confidence with accuracy, and then on mechanisms for persuasion. Moreover, the requirements to overturn an inaccurate majority become greater as the majority becomes

larger—i.e. precisely when the benefits of collective intelligence are most needed. While jury deliberation analyses (Burghardt et al. 2019) suggest that the majority-rules dynamics are largely dominant even when people can express their confidence (i.e. in conversation), it remains possible that some individual juries—overlooked in main effects analysis—did indeed "solve" the crowd classification problem. Nonetheless, such outcomes suggest that a minority overturning a majority is the exception rather than the rule.

The more difficult challenge is that these strategies won't solve the problem for numeric exchange. The irrelevance of social-weighting to Proposition 2 reveals that the crowd classification problem is more of a statistical effect than a reflection of human behavior—thus behavioral solutions are less likely to easily resolve the issue. This difficulty is demonstrated by our analysis of data from experiments (Gürçay et al. 2015) in which subjects shared confidence and other information through natural language discussion, and yet produced comparable outcomes—i.e., consistent with predictions—to experiments without such mechanisms (Becker et al. 2017 and Lorenz et al. 2011).

As the crowd classification problem is a statistical problem, future research may aim for statistical solution. When groups produce numeric estimates, the relative location of key values (median, threshold, truth) determines the effect of social dynamics on binary choice accuracy. For independent beliefs, some tasks have been found to produce regularly structured belief distributions allowing the true value to be inferred from observed properties (Kao et al. 2018). Similarly, it may be possible to identify statistical characteristics that serve as a clue to where in the parameter space a group may find themselves regarding the social dynamics of binary choice accuracy. One strategy may therefore be to "calibrate" decision-making processes to a particular task. By combining theoretical and empirical research as we do here, it may be possible to identify regular properties for a given task of interest.

During the preparation of this manuscript, we explored theoretical strategies for identifying in advance whether social information exchange such as discussion would improve vote accuracy. For numeric estimates, accuracy improvements emerge due to fairly reliable group dynamics. For binary choice votes, however, the dynamics themselves depend on the relative location of initial beliefs and the true value—which is outside the control of the group, and depends entirely on the particulars of the task at hand. The challenge with social influence and binary votes is, as we identify here, that it is simply unreliable from this perspective. At present, we do not know of any way in advance to determine whether social influence will help or harm binary vote accuracy, and that is the heart of the crowd classification problem.

Ultimately, we recommend against voting as a process for the manager concerned with optimizing decision accuracy. A person's binary choice reflect less nuance of belief compared with a numeric estimate. We recommend following a process inspired by the studies referenced in the introduction, which show how group discussion can improve accuracy when carefully structured (Becker et al. 2017). As discussed above, one important factor is collecting independent estimates before people have a chance to talk (Minson et al.

2018). The present study addresses the point of final aggregation, and adds the importance of surveying contributors on their detailed numeric estimates regarding the factor or factors important to the decision at hand, rather than simply taking a vote.

CROWD CLASSIFICATION PROBLEM                                                                                 29Mannes AE, Soll JB, Larrick RP (2014) The wisdom of select crowds. *J. Pers. Soc. Psychol.* 107(2):276.
March JG (1991) How decisions happen in organizations. *Hum.-Comput. Interact.* 6(2):95–117.
Medvec VH, Galinsky AD (2005) Putting more on the table: How making multiple offers can increase the final value of the deal. *HBS Negot. Newsl.* 8:4–6.
Minson JA, Mueller JS, Larrick RP (2018) The Contingent Wisdom of Dyads: When Discussion Enhances vs. Undermines the Accuracy of Collaborative Judgments. *Manag. Sci.* 64(9):4177–4192.
Mossel E, Tamuz O (2017) Opinion exchange dynamics. *Probab. Surv.* 14:155–204.
Navajas J, Niella T, Garbulsky G, Bahrami B, Sigman M (2018) Aggregated knowledge from a small number of debates outperforms the wisdom of large crowds. *Nat. Hum. Behav.* 2:1.
Nofer M, Hinz O (2014) Are crowds on the internet wiser than experts? The case of a stock prediction community. *J. Bus. Econ.* 84(3):303–338.
Page SE (2007) *The difference: How the power of diversity creates better groups, firms, schools, and societies* (Princeton University Press).
Palley AB, Soll JB (2019) Extracting the Wisdom of Crowds When Information Is Shared. *Manag. Sci.*
Parayitam S, Dooley RS (2007) The relationship between conflict and decision outcomes: Moderating effects of cognitive- and affect-based trust in strategic decision-making teams. *Int. J. Confl. Manag. Bowl. Green* 18(1):42–73.
Rivera LA (2012) Hiring as cultural matching: The case of elite professional service firms. *Am. Sociol. Rev.* 77(6):999–1022.
Schnusenberg O (2006) The stock market behaviour prior and subsequent to new highs. *Appl. Financ. Econ.* 16(6):429–438.
Sherif M (1935) A study of some social factors in perception. *Arch. Psychol. Columbia Univ.*
Shore J, Bernstein E, Lazer D (2015) Facts and figuring: An experimental investigation of network structure and performance in information and solution spaces. *Organ. Sci.* 26(5):1432–1446.
Straub VJ, Tsvetkova M, Yasseri T (2020) The cost of coordination can exceed the benefit of collaboration in performing complex tasks. *ArXiv200911038 Nlin Physicsphysics*.
Stroebe W, Nijstad BA, Rietzschel EF (2010) Chapter four-beyond productivity loss in brainstorming groups: The evolution of a question. *Adv. Exp. Soc. Psychol.* 43:157–203.
Surowiecki J (2004) *The wisdom of crowds* (Anchor).
Ven AHVD, Delbecq AL (1974) The effectiveness of nominal, Delphi, and interacting group decision making processes. *Acad. Manage. J.* 17(4):605–621.
Wolf M, Krause J, Carney PA, Bogart A, Kurvers RHJM (2015) Collective intelligence meets medical decision-making: the collective outperforms the best radiologist. *PloS One* 10(8):e0134269.

# E-Companion Appendix

## EC.1. Pre-Registration and Replication Materials.

Pre-registered hypotheses and figures along with all experimental data, analysis code, and simulation code are available for download at: https://osf.io/nj2tz/?view_only=fb08f458331a4d96b54ec4da72a54cb5 . Materials will be posted to Harvard Dataverse for permanent storage upon publication. *PEER REVIEW: because the original pre-registration materials contain author information, this link points to an anonymized but otherwise identical version. The date therefore does not reflect the original pre-registration date, and a URL to the original time-stamped pre-registration will be provided upon publication.*

## EC.2. Appendix for Proposition 1

### EC.2.1. Proofs

*Proposition 1a. When $w_i$ is equal for each agent (they are all equally influential) the majority opinion after any number of timesteps will be the same as the initial opinion, in expectation, even if confidence is correlated with accuracy.*

Proof: The proposition for the case where $F_{maj}=0$ follows from the assumptions, since only the majority can grow, and a consensus is the only absorbing state. This outcome follows because nobody in the majority will ever change their belief, meaning the majority can never shrink; and the number of people switching from minority to majority will always be greater than or equal to zero.

To prove the equilibrium for the case where $F_{maj}>0$, assume without loss of generality that $\Pi>0.5$, i.e. that 1 is the majority opinion. Let $d_{maj}$ be the expected number of people transitioning from the minority to the majority and $d_{min}$ be the expected number of people transitioning majority to minority. And,

$$d_{maj} = \Pi * N * F_{min}$$
$$\text{and } d_{min} = (1-\Pi) * N * F_{maj}.$$

Then if we assume an infinitely large N, the group will be in equilibrium when $d_{maj} = d_{min}$. Rearranging terms, we can state that equation as:

$$\Pi = \frac{F_{min}}{F_{min}+F_{maj}}$$

In very small populations, random chance could lead a minority to grow when $F_{maj}>0$. In this case, the equilibrium would remain the same, as the behavior of those adopting the now-majority opinion will be governed by $F_{maj}$. Although the probability of a minority winning decreases as N increases, we can only say for small N that the social dynamics favor the initial majority in finite time. This is true because as long as the initial $F_{maj}>F_{min}$, the initial majority is likely to grow.

*Proposition 1b. When $w_i$ varies, the unweighted majority opinion will be favored as long as the inequality $\Pi/(1-\Pi) > R$ is satisfied, where $\Pi$ is the proportion holding the initial majority and R is the ratio between the average weight given to minority-belief-holders and the average weight given to majority-belief-holders.*

Proof. A group will favor the simple majority when S>0.5. Note that

$$S = \sum_{i=1}^{N} W_i B_i$$

which simplifies to

$$S = \sum_{i \in B_i = 1} W_i \ .$$

since by assumption $B_i=0$ for all minority belief holders. This summation is also equal to the average weight for all majority belief holders times the number of majority belief holders, i.e.

$$S = \overline{W_{maj}} \Pi N \ .$$

Observe also that

$$\overline{W_{maj}} \Pi N + \overline{W_0}(1-\Pi)N = 1 \tag{1}$$

since the total weight must sum to 1. Then define the ratio R as

$$R = \frac{\overline{W_{min}}}{\overline{W_{maj}}}$$

and rewrite (1) as

$$\overline{W_{maj}} \Pi N + \frac{\overline{W_{maj}}}{\overline{W_{maj}}} \overline{W_{min}}(1-\Pi)N = 1$$

and rearranging and simplifying we can state:

$$\overline{W_{maj}} = \frac{1}{\Pi N + (1-\Pi)NR} \ .$$

Thus the condition S>0.5 i.e. $\overline{W_{maj}} \Pi N > 0.5$ is met when

$$\frac{\Pi N}{\Pi N + (1-\Pi)NR} > 0.5$$

which, rearranging and simplifying, gives the terms stated in the proposition, i.e. that S>0.5 when

$$\frac{\Pi}{(1-\Pi)} > R \ .$$

### EC.2.2. Extended Discussion

Research on numeric belief formation shows that a positive accuracy/stubbornness correlation—more accurate people revise less—can explain the observed benefits of social influence on collective accuracy (Becker et al. 2017, Noriega-Campero et al. 2018). For binary choice, empirical data sometimes shows a positive correlation and sometimes a negative correlation between accuracy and reported confidence, with the magnitude and direction of the correlation depending on the question (Koriat 2012). However, even if this translates into a correlation between accuracy and likelihood of revision, our results will not be affected as long as $F_{min} > F_{maj}$. To see why this is likely, consider the following: a very confident person in the minority may be stubborn, but may yet be more likely to revise than an uncertain person in the majority—as an uncertain person is likely to follow the majority. Thus even if a person is more confident when they are correct, we would still expect $F_{min}$ (uncertain & incorrect) > $F_{min}$ (confident and correct) > $F_{maj}$ (uncertain / incorrect) > $F_{maj}$ (certain/correct). We note that the conditions in Proposition 1b describe how this may change if persuasion is introduced.

One notable deviation from our theoretical prediction is the task in which subjects estimated the total market capitalization of Coca-Cola (blue points, Figure 2): in 3 of 5 trials where the group was initially inaccurate for this question, the majority improved. While this outcome could be explained by sampling error, this task may also represent a deviation from our intended theoretical goal, which is to understand how groups process information, not how groups obtain information. In our initial pre-test, we observed a large variance in responses, indicating that subjects were not able to answer the question via web search. This observation led us to include the question in our experiment. However, the pre-test only provided 60 seconds for answers, while the main experiment provided opportunities to revise and thus gave subjects 3 full minutes. We therefore conclude that this task deviated from our intended empirical context by reflecting information gathering and not just information processing. We note however that our results are nonetheless broadly consistent with our pre-registered predictions, and we offer this post-hoc analysis of this specific task only to illustrate our scope conditions.

### EC.3. Numerical Calculations for Proposition 2

The agent-based model we study here is fully defined by three parameters: $B_0$, the initial belief distribution; A, the adjacency matrix of attention; and R, the number of times each agent updates their belief. The vector of beliefs after R updates is calculated (DeGroot, 1974):

$$B_t = A^R B_0 \ .$$

We present outcomes for belief distributions following a standard normal distribution, and found comparable results for distributions following a log-normal distribution. Code to replicate these simulations for any arbitrary distribution is available in the replication materials.

We must then also define the "vote." For any given threshold T, the proportion Π of people voting "yes" is equal to

$$\Pi = \sum_{i=1}^{N} B_i > T \ .$$

For a given pair of pre- and post-discussion belief distributions, we calculate the pre- and post-discussion vote Π for some threshold T. For a given DeGroot process, we calculate what the vote would be at many different percentiles to ensure robustness. We calculate votes for thresholds T equivalent to every 5$^{th}$ percentile of the standard log-normal distribution.

R code to conduct this numeric calculation is available in an online data repository (see EC.1). We conducted 1,000 numeric calculations for a population of either N=100 or N=1,000 agents embedded in a fully connected (dense) network with individual self-weights drawn from a random uniform [0,1] distribution completing R=10 rounds of revision. Short-term outcomes do not yield consensus, and C therefore represents the post-discussion mean belief, or the "collective belief" in collective intelligence literature (Page 2007).

## EC.4. Extended Methods and Results for Binary Exchange Experiment

Subjects recruited from Amazon Mechanical Turk ("MTurk") completed binary choice estimates before and after observing each others' votes. Estimation tasks consisted of a numeric estimate with a threshold-based binary choice. For example, one question asked subjects to estimate the number of Americans who think that science and technology improves our lives. Subjects were given two response options, "Above 60%" or "Below 60%". Display order was randomized across subjects. Subjects were paid based on the accuracy of their answers (see Fig. A5 for a screenshot of instructions shown to subjects).

A single trial consisted of 20 subjects simultaneously responding to a single question. After first providing an independent estimate, subjects were shown a list of other subjects' responses and a summary tally of votes, and were given the opportunity to provide a new estimate (see screenshot in Fig. A5). Subjects were then shown the revised responses of other subjects and given another opportunity to revise. Subjects provided three estimates in total—one independent estimate and two socially influenced estimates, following prior laboratory experiments on belief accuracy (Becker et al. 2017, Gürçay et al. 2015, Lorenz et al. 2011). Display order for social information was randomized across subjects.

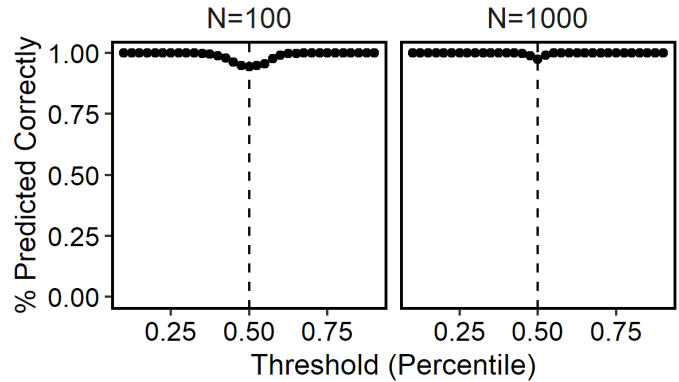

**Figure A1. Fit of approximation for proposition 2c.**

*Notes: Points show the proportion of numeric experiments correctly predicted by Proposition 2c for two population sizes.*

Each subject answered only one question (i.e. participated in only one trial). Each group/trial was assigned one of five possible estimation tasks: the count of candies in a jar, the calories in a dessert, employment statistics, public attitudes toward technology, and the market capitalization of Coca-Cola. For each task, each group/trial was assigned one of three unique numeric thresholds. In total, this produced 15 unique questions. Prior to conducting our experiment, thresholds were selected by collecting 150 numeric responses to each question (these subjects were excluded from further participation). We chose thresholds based on an analysis of the response distributions. Our goal was to target specific accuracy levels by setting thresholds at precise numeric percentiles—e.g., setting the threshold at the $40^{th}$ percentile response with the intent of obtaining a 40% accuracy rate. However, we found during pilot testing that this process was extremely imprecise, due most likely to compounded sampling error. Thus, we selected thresholds based on subjective examination of the response distribution, seeking three broad ranges for each question: majority correct, majority incorrect, and even split.

An additional pre-registered analysis uses OLS to test the hypothesis that the magnitude of the change in accuracy (measured as the absolute value) is predicted by the square of the size of the initial majority (absolute value of 50 minus the initial accuracy). This analysis is statistically significant ($P<0.01$).

**EC.4.1. Exact wording and threshold for questions**

- The jar in this image contains nothing but standard M&M's. How many M&M's are in the jar? (Thresholds: 275, 350, 500. Answer: 797.)

- A 2014 survey asked Americans whether science and technology make our lives better (easier, healthier, more comfortable). What percentage of respondents agreed that science and technology are making our lives better? (Thresholds: 60%, 70%, 80%. Answer: 80.5%)
- How many calories do you think are in the yogurt dessert pictured here? (Thresholds: 175, 200, 250. Answer: 180)
- As of Friday November 16th, the stock market valued PEPSI CO at $166.35 billion. How much is rival company COCA COLA worth? (Thresholds: $250 bn, $300 bn, 350 bn. Answer: $216 bn)
- According to a 2016 survey of Americans, the most common places to be employed are construction, education, and healthcare. What percentage of employed Americans work in one of these three industries?" (Thresholds: 30%, 40%, 55%. Answer: 28.6%)

**EC.4.2 Comparison of Manuscript Results and Pre-Registered Hypotheses**

We present the results of a pre-registered analysis in support of Proposition 1. All pre-registered tests are statistically significant at $P<0.01$ level except as follows. Pilot Analysis 1 indicates that the effect of initial accuracy is significant at $P<0.05$, and Pilot Analysis 2 indicates that the effect of initial accuracy is significant at $P<0.09$ (as reported in the main text). For the main experiment, Analysis 3 is significant at $P<0.02$ for groups that are initially inaccurate.

During the peer review process, we discovered a coding error in the script used to generate the pre-registered version of Figure 2. We present a corrected version in this manuscript, and the initial pre-registered version remains available with the pre-registration materials. Our replication materials include the original erroneous script used to produce the pre-registration materials as well as the corrected version.

**EC4.3. Model Fit**

To evaluate the fit for our theoretical model, we compare $R^2$ for four models: (1) a naïve linear model regressing change on whether the initial majority was accurate; (2) our pre-registered theoretical model as shown in Figure 2; (3) our pre-registered theoretical model with accuracy/revision correlation as shown in Figure A3; (4) and an empirically fitted polynomial curve matching the shape of the theoretical curve as shown in Figure 2. $R^2$ for these models are respectively: 0.41, 0.25, 0.35, and 0.43. While the goal in theoretical modeling is to capture broad qualitative characteristics of a social system (Macy and Willer 2002), and not typically to predict effect sizes, the fit of each model provides a broad illustration of the relative contribution of different theoretical components.

The generally good fit of the naïve linear model reflects the qualitative theoretical prediction that dynamics bifurcate into two broad categories based on whether the initial vote was accurate or inaccurate. Our pre-registered theoretical model produces a comparatively poorer fit, reflecting both the high variance in empirical data as well as the fact that we calibrated our model to only one question but tested five. The inclusion of an accuracy/adjustment correlation moderately improves outcomes, indicating the theoretical importance of this relationship. Finally, we note that the polynomial curve produces only a nominally increased performance compared with the naïve linear model, suggesting that the accurate/inaccurate distinction tested by the naïve linear model is the most important predictor. (While a polynomial curve can be intentionally overfit to artificially inflate $R^2$, such an overfitted would produce a many-peaked "squiggle" whereas our tested curve fits the theoretically predicted shape.)

## EC.5. Methods and Results for Pilot Experiment

### EC.5.1. Methods

We used a web-based laboratory experiment with the Empirica.ly platform (Almaatouq, Becker, et al. 2020) Subjects recruited from Amazon Mechanical Turk ("MTurk") completed binary choice estimates before and after observing simulated social information. The estimation task asked subjects to estimate the number of M&M candies in a jar based on a photograph. Subjects provided their responses by selecting one of two options ("Above 350" and "Below 350") provided in random order. This threshold was selected based on pre-test data in which subjects were simply asked to provide a numeric response; the median estimate was approximately 350. We chose the median as the basis for the option set under the assumption that it would yield approximately 50% accuracy in a binary choice estimate.

After first providing an independent estimate, subjects were shown the estimates of 10 digital confederates (simulated peers) and prompted to provide their estimate again, i.e. given the opportunity to revise their initial estimate. Subjects were then shown responses by the digital confederates again (which did not change) and given the opportunity to revise their estimate again. By this method, subjects provided three estimates in total—one independent estimate and two socially influenced estimates, following the method described in previous research on numeric estimates (Becker et al. 2017). Subjects were randomly assigned to one of nine conditions, where a condition determined the amount of (dis)agreement shown by the digital confederates, ranging from 10% (1 of 10 digital confederates agreed with the subject) to 90% (9 of 10 digital confederates agreed with the subject). We collected responses by 50 subjects per condition, generating data for 450 subjects in total, following our pre-registered sample size.

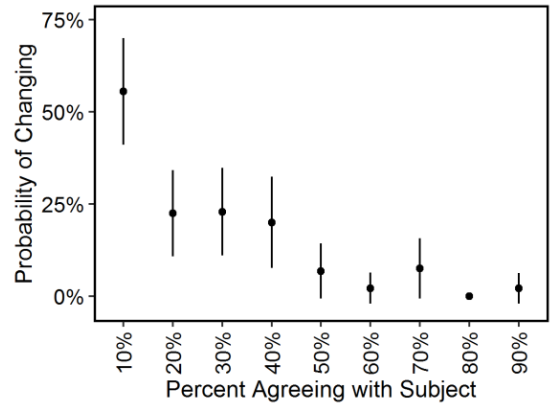

**Figure A2. Probability of Revising in Individual Experiment**

*Notes: Error bars show 95% confidence intervals.*

To ensure that the simulated social information provided by digital confederates was realistic, the procedure for our pilot study followed identically the procedure commonly used for synchronized web-based network experiments, including the procedure planned for our main study. Thus, although subjects completed the estimation tasks individually, they were nonetheless required to arrive at the experimental website simultaneously and wait for other players to arrive. Except for experimentally controlled variation, the subject experience for our pilot study was identical to the subject experience for our main study.

To facilitate simultaneous participation, subjects were sent a notification prior to the scheduled experiment describing the research. Subjects were told the task would take a maximum of three minutes and offered a guaranteed pay of $0.50 for participating. To incentivize honest reporting, subjects were offered a bonus up to $0.50 based on accuracy, for a total maximum payment of $1.00.

Subjects were recruited from a panel of 21,121 MTurk members who had previously enrolled to participate in research with our university research lab. Invitations for the pilot study were sent to a random sample of 1,000 subjects from this panel. Upon arrival at the study website, subjects provided informed consent, read instructions describing the task, and provided a "username." Upon completion of this introductory content, subjects were shown a countdown timer until the task began.

**EC.5.2. Results**

To estimate the effect of social information on individual beliefs, we examine the change in answers from Round 1 (independent estimate) to Round 3 (influenced by digital confederates). As shown in Figure A2, subjects were more likely ($P<0.001$, logistic regression controlling for initial accuracy) to change their answer when the majority of confederates disagreed (30% of subjects switched their answer) than when the majority of subjects agreed (only 4% switched their answer). Subjects were also unlikely to change their answer when half of the digital confederates agreed (only 7% switched). Consistent with research on conformity (Asch 1951), we observe a sharp drop-off in the effect of social influence as the level of apparent agreement increased. While over 50% of subjects revised their answer when 9/10 confederates disagreed with the subject, fewer than 25% of all remaining conditions revised their answer. Figure 1 shows the primary result of interest, which is an estimation of individual behavior as a function of social information, $P(x)$ where x is the proportion of social agreement and $P(x)$ is the probability that a subject will revise their answer. This function is a sufficient statistic for calibrating our theoretical model.

### EC.5.3. Empirically calibrated model of group behavior

Our empirically calibrated simulations were run in R according to the model described in the main text and as follows:

- Initial conditions at time for a population of N agents are determined by a vector of initial beliefs $B_{t=0}$ and by $P(B)$, the probability of revision as a function of percentage of peer agreement (i.e. Figure A2).
- At each timestep $t+1$, agents all determine whether to flip their belief with probability $P(B_t)$.
- $P(B)$ may account for whether an agent holds the correct belief, i.e. generating a potential revision/accuracy correlation.
- $P(B)$ is estimated from the data shown in Figure A2.
- We simulate 2 rounds of revision for a population of N=20, consistent with our experimental conditions.

The results of this calibrated simulation corroborated our initial theoretical results. These results also allowed us to generate an empirically calibrated prediction for the effect of social influence at different levels of disagreement, which shows a non-monotonic effect of disagreement, as shown in Figure 2. We find that the change in group belief is smallest when the group is near full agreement and when the group is split 50/50 or close to an even split: in both cases, very few individuals are likely to change their answer. Our initial pre-registered prediction was based off a parsimonious version of the model that did not account for the accuracy/revision adjustment. The dashed line in Figure A3 shows a version of the model that does account for the accuracy/revision adjustment, which became more clearly statistically significant after additional data collection.

### EC.6. Supplementary Figures and Table.

Provided at the end of this Appendix.

*Outcome: % trials consistent with model.*

| Dataset | Smallest Increment | | |
|---|---|---|---|
| | 0.01 | 0.05 | 0.10 |
| Becker et al. | 84% | 82% | 82% |
| Gurcay et al. | 76% | 74% | 74% |
| Lorenz et al. | 79% | 79% | 78% |

**Table A1. Proposition 2 model fit as a function of threshold resolution.**
*Notes. Empirical calculations depend on measuring outcomes at each of a discrete set of thresholds. Our main results report thresholds in increments of 0.01, and this table also shows outcomes for increments of 0.05 and 0.10. Thresholds correspond to quantiles as described in the main text.*

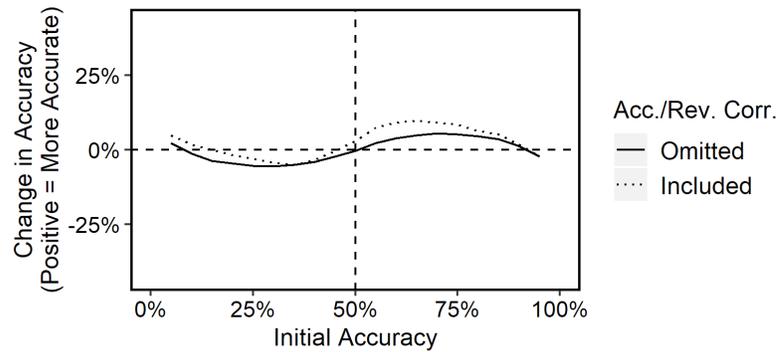

**Figure A3. Effect of accuracy/revision correlation on empirically calibrated model.**
*Notes: solid line shows main effect as reported in main text. Dashed line shows model with accuracy/revision correlation.*

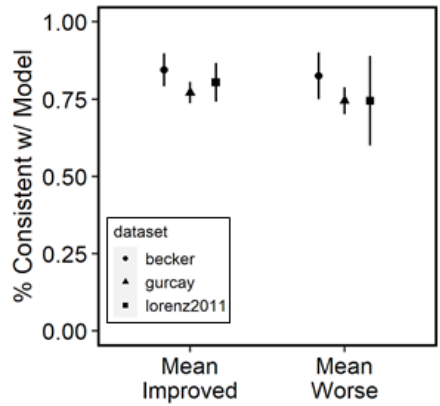

**Figure A4. Fit of Proposition 2 as a function of initial accuracy.**
*Notes: This figure demonstrates that the fit of Proposition 2 fits does not depend on initial accuracy.*

## Instructions

### Game play:
- You will be asked to estimate a numeric quantity. You will have 45 seconds to answer.
- After giving your answer, you will get two chances to revise your estimate.
- You may see information on other people's estimates.
- This task will take a maximum of 5 minutes.

### Payment (via bonus to MTurk #AX00000000):
- You get a guaranteed minimum of 50 cents for participating. *Double your earnings with accurate estimates!*
- Earn an extra $0.15 for the first response, $0.15 for the second response, and $0.20 for the third response.
- **Total maximum payment is $1.00. Minimum payment is $0.50. ($6-12/hour)**

[Next »]

**Figure A5. Screenshot of the instructions**
*Notes: Estimated game time and hourly earnings include estimated maximum waiting time.*

**Figure A6. Screenshot of "Round 2" of the experimental interface.**
*Notes: Prior to beginning the experiment, subjects provided a "username" that is associated with their responses as shown in the list on the right. If a subject failed to enter a username, it is indicated as 'NA'. If a subject failed to enter a response, it is indicated as 'No Response' and not included in the summary.*